\documentclass[final,5p,times,twocolumn]{elsarticle} 
\usepackage{amssymb}

\usepackage[table,dvipsnames,x11names]{xcolor}
\usepackage[table,dvipsnames,x11names,addedmarkup=bf,addedcolor=BrickRed,deletedmarkup=sout,deletedcolor=Blue4]{changes}
\usepackage{amsmath}
\usepackage{multirow}
\usepackage{booktabs}
\usepackage{bm}
\usepackage{braket}
\usepackage{amsmath}
\usepackage{graphicx}
\usepackage[normalem]{ulem}
\biboptions{sort&compress}

\graphicspath{{./}}

\usepackage[pdfstartview=FitH, colorlinks,
 linkcolor={red!60!black!},
 anchorcolor={green!80!black!},
 urlcolor={blue!80!black!},
 citecolor={blue!50!black!}]{hyperref}

\journal{Physics Letters B}

\begin{document}

\begin{frontmatter}

\title{Model-independent mass determination of near-threshold states\\from short-range production}

\author[ad1,ad2]{Yong-Hui Lin}
\author[ad2,ad3]{Hans-Werner~Hammer}
\author[ad4,ad5,ad1]{Ulf-G.~Mei{\ss}ner}

\address[ad1]{Peng Huanwu Collaborative Center for Research and Education, International Institute for Interdisciplinary and Frontiers,\\ Beihang University, Beijing 100191, China}

\address[ad2]{Institut für Kernphysik, Technische Universität Darmstadt, 
    64289 Darmstadt, Germany}

\address[ad3]{ExtreMe Matter Institute EMMI and Helmholtz Forschungsakademie Hessen f\"ur FAIR (HFHF),\\ GSI Helmholtzzentrum 
    für Schwerionenforschung GmbH, 64291 Darmstadt, Germany}

\address[ad4]{Helmholtz--Institut f\"ur Strahlen- und Kernphysik (Theorie)\\ 
    and Bethe Center for Theoretical Physics, Universit\"at Bonn, D-53115 Bonn, Germany}

\address[ad5]{Institute for Advanced
    Simulation (IAS-4), \\
    Forschungszentrum J\"ulich, D-52425  J\"ulich, Germany}

\begin{abstract}
We propose a novel observable for the precision measurements of a wide class of near-threshold dimer states:
the short-range production rate of a dimer--spectator two-body system, composed of the given near-threshold state and one of its constituents. 
Within the framework of nonrelativistic effective field theory, these production rates exhibit characteristic line shapes for the specific partial wave and reach a model-independent minimum. 
This feature enables a precise extraction of their masses from experimental data, 
provided that the line shape can be resolved with sufficient accuracy. 
Applying this novel method to both the $T_{b\bar{b}1}(10610)B$ and $T_{b\bar{b}1}(10650)B^*$ systems allows for a precise determination of the binding energy $\delta$ of the $T_{b\bar{b}1}(10610)$ and $T_{b\bar{b}1}(10650)$ via the relation of $\delta=-{E_{\text{dip}}^{\text{exp}}}/{0.1983}$ once the respective dip position $E_{\text{dip}}^{\text{exp}}$ is experimentally identified.
\end{abstract}


\end{frontmatter}

\section{Introduction}
A precise determination of the masses of near-threshold states is essential for unveiling their underlying structure. 
The binding energy $\delta$, defined as the mass difference between such a near-threshold state with mass $m$ and its nearby threshold defined by the sum of the masses of its constituents $m_i+m_j$, i.e., $\delta=m_i+m_j-m$, 
encodes key information about the low-energy interactions of the particles involved. 
The presence of a near-threshold two-body state (called dimer) indicates an attractive interaction between the two constituent particles, suggesting the formation of a loosely bound system. 
Such behavior is a necessary condition for interpreting the near-threshold state as a hadronic molecule, a configuration that has emerged as a compelling alternative to conventional quark-antiquark mesons and three-quark baryons in describing exotic hadrons, see e.g. Refs.~\cite{Hosaka:2016pey,Esposito:2016noz,Guo:2017jvc,Olsen:2017bmm,Karliner:2017qhf,Kalashnikova:2018vkv,Brambilla:2019esw,Meng:2022ozq,Liu:2024uxn,Chen:2024eaq} for recent reviews.

In scattering theory, the mass of a physical state is defined as the pole position in the corresponding $S$-matrix, a quantity that cannot be directly accessed in experiments. 
This leads often to a troublesome model dependence when extracting the mass from experimental observables, such as differential cross-section line shapes. 
For further details, see the section titled ``Resonances'' in the Review of Particle Physics (RPP)~\cite{ParticleDataGroup:2024cfk}.
In this work, we propose a novel observable, the point production rates of a two-body system consisting of the given near-threshold dimer state and one of its constituents.
This quantity is accessible in short-range production of the dimer--spectator system and enables a model-independent extraction of the binding energy from experimental data. Note that the approach proposed  here is very different in nature from the recently proposed method to
precisely pin down the mass of the $X(3872)$ by measuring the $X(3872)\gamma$ line shape~\cite{Guo:2019qcn}.

Our method relies on the fact that three-body systems of a near-threshold dimer state and one of its constituents have a large two-body scattering length.
Such systems in turn display universal behavior related to an approximate non-relativistic conformal symmetry for low energies small compared to the energy scale set by the range of their interactions~\cite{Mehen:1999nd,Braaten:2004rn,Nishida:2007pj}. 
This symmetry strongly constrains the behavior of few-body systems with small relative momenta. The system formed by a near-threshold dimer state and one of its constituents constitutes an ideal physical setup whose dynamics is governed by the symmetry.
It exhibits universal behavior characterized solely by two intrinsic properties, their mass and quantum numbers~\cite{Hammer:2021zxb,Braaten:2021iot,Braaten:2023acw}. As a result, one can construct observables that contain information about the mass of the near-threshold dimer state. 

In the following, we demonstrate that the point production rate of the dimer-spectator system constitutes such an observable which can be used to extract the mass of the near-threshold dimer state. 

\section{Dimer--spectator dynamics in NREFT}
We work within the dimer--spectator framework, treating the near-threshold state as a composite dimer field formed by two threshold particles with masses $(M_l, M_h)$, where $M_l \le M_h$ and the mass ratio is defined as $r \equiv M_l / M_h \in (0, 1]$. The reduced mass of this two-body system is then
\begin{equation}
    \mu=\frac{M_l M_h}{M_l+M_h}=\frac{M_h r}{1+r}=\frac{M_l}{1+r}.
\end{equation}
The dimer interacts with a third spectator particle, which, without loss of generality, is taken to be the light particle $M_l$.

\begin{figure}[htb]
    \begin{center}
        \includegraphics[width=0.45\textwidth]{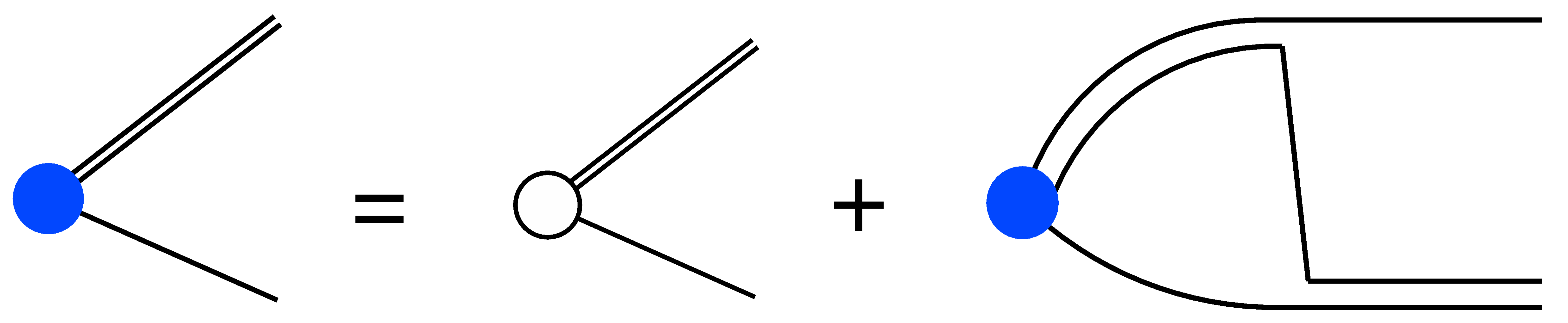}
    \end{center}
    \caption{\label{fig: pointppd}{Diagram for the point production of the dimer-spectator pair. }}
\end{figure}
A nonrelativistic effective field theory (NREFT) can be formulated to describe the dynamics of such near-threshold dimer--spectator systems, in which only short-range contact interactions between the dimer and the constituent particles are retained explicitly at leading order (LO).
This treatment is applicable if the relevant energies are low enough such that any meson exchanges are not resolved explicitly.
In the single channel case, the partial-wave short-range production amplitude $\Gamma_L$ of the dimer--spectator system
satisfies the integral equation shown diagrammatically in Fig.~\ref{fig: pointppd},
\begin{align}\label{eq: inteq0}
    &\Gamma_L(E,p)= A_L(E,p)+ (-1)^L C_{SI}\int_0^\infty \frac{dq\, q^2}{\pi \mu}\frac{M_h}{p q} \notag\\
    &\phantom{xxx} \times\frac{\Gamma_L(E,q)Q_L\left(\frac{-2\mu E+q^2+p^2}{2\mu p q/M_h}\right)}{-1/a+\sqrt{-2\mu E+\left(\mu/\tilde{\mu}\right) q^2 -i \epsilon}}\,.
\end{align}
{Here $Q_L$ is the Legendre polynomial of the second kind,
    \begin{equation}
        Q_L(z-i\varepsilon)=\frac{(-1)^L}{2}\int_{-1}^1 dx \frac{P_L(x)}{x+z-i\varepsilon}.
\end{equation}}
The quantity $\tilde{\mu}^{-1}=(M_l+M_h)^{-1}+M_l^{-1}=(\frac1{r+1}+\frac1r) M_h^{-1}$ denotes the reduced mass of the dimer--spectator system. 
Furthermore, $a$ denotes the scattering length of the two-body threshold system, with
\begin{equation}
    a=1/\gamma={\rm sgn}(\delta)/\sqrt{2\mu|\delta|},
\end{equation}
which is positive for a bound state with the positive binding energy $\delta$, and where $\gamma$ denotes the corresponding binding momentum. 
$C_{SI}\equiv {\langle{\cal O}\rangle_{SI}} S_3/{S_1}$ is a quantum-number dependent factor,
where the symmetry factor $S_1$ accounts for identical particle contributions in the self-energy of the dimer: $S_1=2$ if the dimer consists of two identical constituents, and $S_1=1$ otherwise. 
{$S_3$ denotes the symmetry factor associated with the potential kernel appearing in the homogeneous term on the right-hand side of the integral equation shown in Fig.~\ref{fig: pointppd}, which is parameterized at leading order in NREFT by a one-particle-exchange tree diagram. Details of the construction of $S_3$
    can be found in Ref.~\cite{Wilbring:2016bda}.}
$\langle{\cal O}\rangle_{SI}$ represents the normalized partial-wave projected prefactor of the dimer--spectator scattering kernel in the integral equation, for the given spin $S$ and isospin $I$.

The partial-wave projected bare point production amplitude is taken in the general form $A_L = g_L p^L$, where $L=0,1,2,...\,$ denotes the relevant orbital angular momentum\footnote{Here, we take only the minimal momentum dependence of the source into account. In principle, there could be higher-order momentum-dependent terms that account for corrections to a point source.}.
Implementing the variable transformation 
\begin{align}
    E\to|\delta|(x-1),\quad q\to \sqrt{2\tilde{\mu}|\delta|}\sqrt{y}\,,
\end{align}
the integral equation~\eqref{eq: inteq0} can be rewritten as
\begin{align}\label{eq: ppdint}
    &\Gamma_L(x,z)=g_L\left(2 \tilde{\mu} |\delta|\right)^{L/2}z^{L/2}+ (-1)^L C_{SI}\int_0^\infty \frac{dy}{2\pi\sqrt{z}} \notag\\
    &\phantom{xxx} \times\frac{(1+r)^2}{r\sqrt{1+2r}}\frac{Q_L\left(\frac{(1+r)^2(y+z)-(1+2r)(x-1)}{2r(1+r) \sqrt{y z}}\right)}{-{\rm sgn}(\delta)+\sqrt{1-x+y -i \epsilon}}\,\Gamma_L(x,y)\,.
\end{align}
{If the heavy particle is taken as the spectator, the corresponding integral equation reads
    \begin{align}\label{eq: ppdint_h}
        &\Gamma_L(x,z)=g_L\left(2 \tilde{\mu} |\delta|\right)^{L/2}z^{L/2}+ (-1)^L C_{SI}\int_0^\infty \frac{dy}{2\pi\sqrt{z}} \notag\\
        &\phantom{xxx} \times\frac{(1+r)^2}{\sqrt{r(2+r)}}\frac{Q_L\left(\frac{(1+r)^2(y+z)-r(2+r)(x-1)}{2(1+r) \sqrt{y z}}\right)}{-{\rm sgn}(\delta)+\sqrt{1-x+y -i \epsilon}}\,\Gamma_L(x,y)\,.
    \end{align}
As the mass ratio $r$ approaches 1, the integral equations for the dimer--light-spectator [Eq.~\eqref{eq: ppdint}] and dimer--heavy-spectator [Eq.~\eqref{eq: ppdint_h}] configurations become identical. In the following, we mainly focus on the systems with mass ratio $r$ close to 1, for which the marginal effects induced by the $r$-dependent coefficients are negligible.  
Moreover, for a virtual dimer the above integral equations remain valid, differing only by the sign of the scattering length $a$. Since a virtual state can appear only as an intermediate state, the associated large-scattering-length dynamics can be probed only in the three-body final state, which is beyond the scope of the present study.}
The one-shell total energy $E$ of the dimer--spectator system with a bound-state dimer 
in the center-of-mass frame is given by
\begin{equation}
    E=\frac{p^2}{2\tilde{\mu}}-\frac{1}{2\mu a^2}=\frac{p^2}{2\tilde{\mu}}-|\delta|\,.
\end{equation}

The point-production rate of such dimer--spectator system is then given by
\begin{equation}\label{eq: rate}
    R(x)=\int_{-1}^{+1} d \hat{z} \, \frac{\tilde{\mu}\sqrt{\tilde{\mu}|\delta|}}{\sqrt{2}\pi} \left|\sum_L(2L+1)P_L(\hat{z})\Gamma_{L}(x)\right|^2\sqrt{x},
\end{equation}
where $\Gamma_{L}(x)\equiv \Gamma_{L}(x,x)$ denotes the on-shell value of the point production amplitude
in Eq.~\eqref{eq: ppdint}, corresponding to the case $z=x$.

\section{Point production of $Z_b B$ and $Z_b^\prime B^*$ and corrections}
From Eqs.~\eqref{eq: ppdint} and \eqref{eq: rate}, it is evident that the point production rate is entirely determined by the bare coupling constant $g_L$, the dimer--spectator reduced mass $\tilde{\mu}$, the corresponding mass ratio $r$, the binding energy $\delta$ of the near-threshold dimer, and the quantum number factor $C_{SI}$.
We now present a numerical analysis of how these parameters affect the point production rate. Notably, $g_L$, $\tilde{\mu}$, and $\delta$ enter the expression as overall scale factors, meaning they do not influence the line shape of the point production rate. Therefore, we fix $g_L=1$ for simplicity. 
For illustration, we focus on the $Z_b B$ and $Z_b^\prime B^*$ systems, aka $T_{b\bar{b}1}(10610)B$ and $T_{b\bar{b}1}(10650)B^*$, as starting points for our discussion. The scattering properties of these systems were already considered in \cite{Lin:2017dbo}.
We then vary the parameters $r$ and $C_{SI}$ 
to explore their roles on the shape of the  point production rate. 

\begin{figure}[htb]
    \begin{center}
        \includegraphics[width=0.48\textwidth]{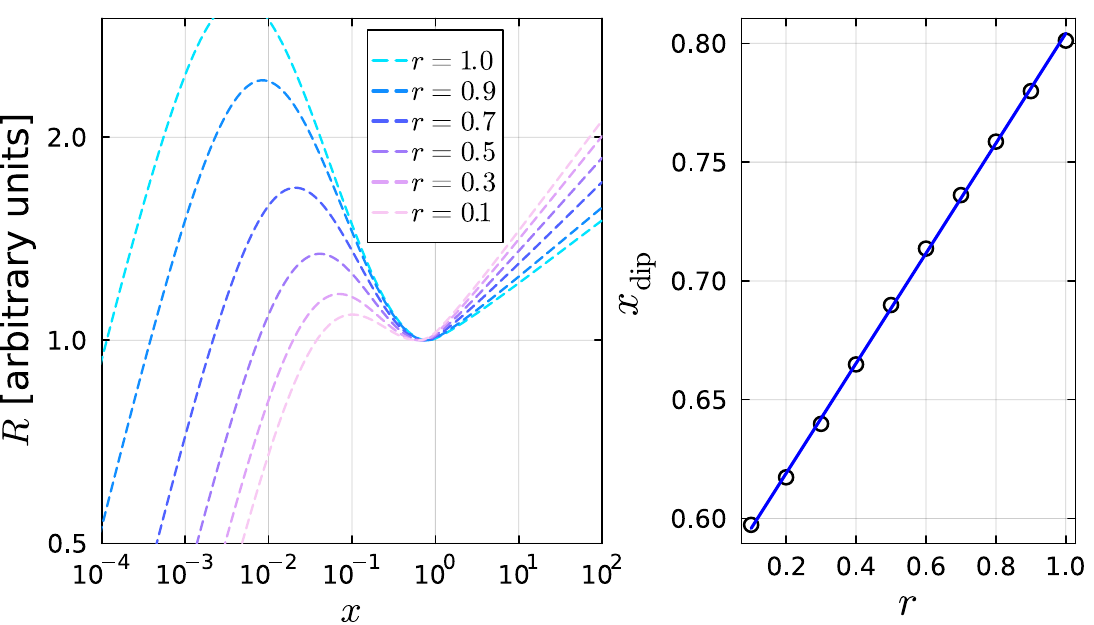}
    \end{center}
    \caption{\label{fig: rvary}{Variation of the $S$-wave point production rate with the mass ratio $r$ (left) for the $Z_b B$ system with $(S, I)=(1,3/2)$ and the position $x_{\rm dip}$ of the characteristic dip as an approximately linear function of $r$ (right). Since the undetermined constant $g_0$ in the bare production amplitude is set to 1 in the plots, the resulting short-range production rate is presented in arbitrary units.}}
\end{figure}

The relevant masses in the latest version of the RPP are given by~\cite{ParticleDataGroup:2024cfk}
\begin{align}\label{eq: masses}
    &M_B\equiv M_{B^{+}}=5279.41(7)\,{\rm MeV},\: M_B^*=5324.75(20)\,{\rm MeV}\,,\notag\\
    &M_Z\equiv M_{T_{b\bar{b}1}(10610)^+}=10607.2(20)\,{\rm MeV}\,, \notag\\
    & M_{Z_b^\prime}\equiv M_{T_{b\bar{b}1}(10650)^+}=10652.2(15)\,{\rm MeV}\,.
\end{align}
Thus the mass ratio is given by $r = 0.99148(4)$ for the $Z_b B$ system, and $r = 1$ for $Z_b^\prime B^*$.
The $S$-wave point production rates for the $Z_b B$ channel with quantum number $(S,I)=(1,3/2)$, corresponding to $C_{1\frac32}=1/2$, are presented in Fig.~\ref{fig: rvary}. Here, $\delta_{Z_b}=1\,{\rm MeV}$ is used for illustration.

As shown in the left panel, 
the $S$-wave production rate rises from zero at threshold ($E=-\delta$ or $x=0$) to a maximum near $E_p=1/(2\tilde{\mu} \tilde{a}^2)$, where $\tilde{a}$ denotes the dimer--spectator scattering length. The latter can be extracted from elastic  dimer--spectator scattering within the same NREFT framework, as discussed in Ref.~\cite{Lin:2017dbo}).
{The rate then decreases to a local minimum at 
$E=0$ ($x=1$), where the relative momentum between the dimer and the spectator ($\sqrt{2\tilde{\mu} |\delta|}$) becomes comparable to the dimer binding momentum ($\sqrt{2{\mu} |\delta|}$), giving rise to a characteristic peak-dip structure that is universal for channels with the same $C_{IS}$. This dip originates from a strong suppression of the on-shell dimer-particle production amplitude $\Gamma(x)$, caused by the breakup of the dimer state, i.e., the opening of a genuine inelastic three-body threshold. Accordingly, the local minimum of $R(x)$ near $x=1$ can be interpreted as a spectroscopic signature of the dimer breakup threshold.}

Note that the rates are normalized to unity at the dip position for each value of $r$ in the figure.
In particular, the right panel shows that the dip position follows an approximately linear dependence on the mass ratio. A fit to the numerical data points gives $x_{\rm dip}=0.2308r+0.5729$. 
This correlation enables an experimental determination of the dimer binding energy by locating the dip in the point production rate of the associated dimer--spectator system.

\begin{figure}[htb]
    \begin{center}
        \includegraphics[width=0.48\textwidth]{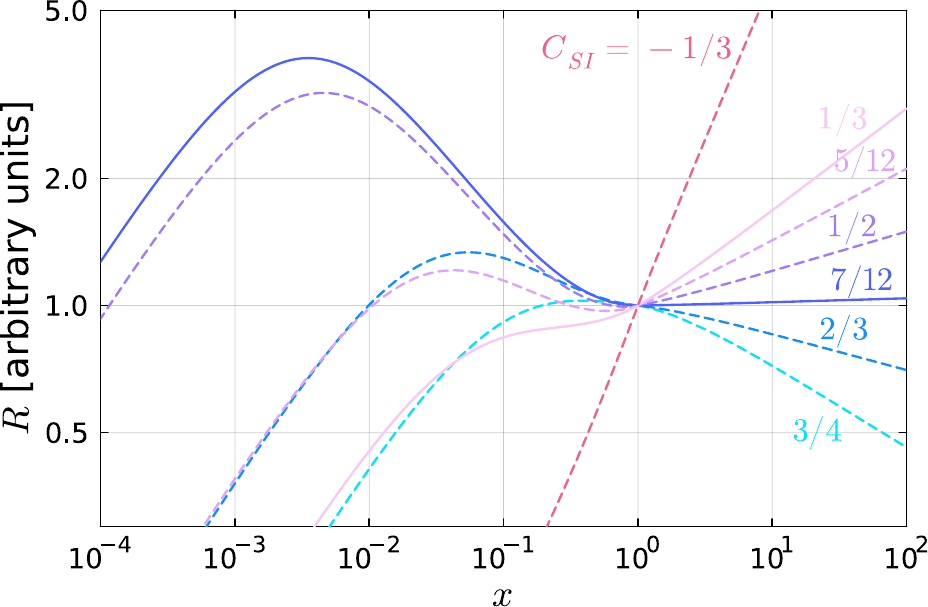}
    \end{center}
    \caption{\label{fig: cvary}{Variation of the $S$-wave point production rate with the quantum number factor $C_{SI}$.
    The characteristic peak-dip structure only appears when $4/12<C_{SI}<7/12$ for $r=1$.}}
\end{figure}

Next, we consider the effect of the quantum number factor $C_{SI}$ by fixing $r=1$. 
The variation of the $S$-wave point production rate with $C_{SI}$ is shown in Fig.~\ref{fig: cvary}, where all rates are normalized to unity at $x = 1$ to facilitate comparison across different $C_{SI}$ values.
It is found that the interesting peak--dip structure emerges only within a narrow window of $C_{SI}$. 
For $r=1$, this range is identified as $4/12 < C_{SI} < 7/12$. 
The range exhibits a slight dependence on the mass ratio $r$: when $r$ is reduced to 0.5, the peak--dip region shifts to $5/12 < C_{SI} < 5/8$, as revealed by numerical calculations. 

Our calculations demonstrate that the point production process of a dimer--spectator system offers a model-independent method to determine the binding energy of a near-threshold dimer state.
For a given near-threshold state that can be treated as a bound state of the corresponding threshold particles with positive binding energy $\delta$, there exists a specific partial wave of the dimer--spectator two-body system in which the point production rate exhibits a characteristic dip.
The position of this dip is located at $\delta(x_{\rm dip} - 1)$, 
where $x_{\rm dip}$ is a universal, model-independent constant.
{As discussed in Ref.~\cite{Fu:2025joa}, in the coupled-channel case the relevant critical factors correspond to the eigenvalues of the partial-wave-factor matrix, which differ from the single-channel value. Nevertheless, if these eigenvalues still lie within the same window, the characteristic line shape persists and the dip position remains unchanged. A detailed extension to the coupled channel problem is left for future work.} 

Next, we consider the effect of a finite effective range $\rho$ on the universal relation between 
$x_\mathrm{dip}$ and $\delta$. Due to space constraints, we summarize the key results and relegate a more detailed discussion of effective-range corrections to the~\ref{app:ERC}.
We find that the direct determination of the binding energy of a near-threshold state from the dip position is valid only within a narrow window near the zero-range limit, i.e. for $-0.1 \lesssim \gamma\rho\lesssim 0.1$. For effective ranges in this range, the linear relation between $x_\mathrm{dip}$ and $\delta$ persists.
For larger effective ranges, a simple determination of the dip position is no longer sufficient. In this case, a multi-parameter analysis of the line shape of the short-range production rate is required to determine both the binding energy of the near-threshold dimer state and the effective range of the corresponding two-body subsystem. However, using short-range effective field theory the dependence of the line shape on the effective range can be readily calculated if $|\gamma\rho| \lesssim 0.3$. If the effective range is even larger the state moves away from the threshold and our method is not applicable.

In some systems, three-body forces enter already at LO in short-range effective field theory, depending on the spin-isospin channel, thereby inducing corrections to the integral equation in Eq.~\eqref{eq: ppdint}. The three-nucleon system can serve as a guide for our intuition.
In the spin-doublet channel of neutron-deuteron scattering, a three-body force is required for renormalization at leading order~\cite{Bedaque:1998kg,Bedaque:1998km,Bedaque:1999ve}, naturally giving rise to the triton as an Efimov state~\cite{Efimov:1970zz}.
In contrast, in the spin-quartet channel, where no three-body bound states are present, three-body forces are strongly suppressed~\cite{Bedaque:1997qi,Bedaque:1998mb}. In this case, $S$-wave three-body forces are forbidden due to the Pauli exclusion principle.
The emergence of three-body bound states arising from the Efimov effect in short-range EFT is therefore directly linked to the necessity of three-body forces at leading order~\cite{Hammer:2010kp}. 
Note, however, that this reasoning does not apply to relativistic formulations of the three-body problem~\cite{Epelbaum:2016ffd}.
Whether three-body forces play a significant role in a given dimer--spectator system can thus be inferred phenomenologically by examining the existence of three-body Efimov states, as explored for the $Z_b B$ and $Z_b^\prime B^*$ systems in Ref.~\cite{Lin:2017dbo}.
In the absence of three-body bound states, the binding energy of the near-threshold dimer can be extracted model-independently from its experimentally measured point production rate based on our method.

\section{Results for the $Z_b B$, $Z_b^\prime B^*$, and $XD$ cases}
Finally, we present theoretical predictions for the total point production rate, including both $S$- and $P$-wave contributions, for the $Z_b B$ and $Z_b^\prime B^*$ systems, assuming that the $Z_b$ and $Z_b^\prime$ are bound states with positive binding energies.
The RPP masses given in Eq.~\eqref{eq: masses} yields binding energies of $\delta_{Z_b}=-3.0(20)\,{\rm MeV}$ and $\delta_{Z_b^\prime}=-2.7(32)\,{\rm MeV}$ for the $Z_b$ and $Z_b^\prime$, respectively, both with substantial uncertainties.
Although the central values reported in the RPP remain negative based on current measurements,
a wide range of theoretical studies support the existence of bound states in the $B^*\bar{B}$-$B\bar{B}^*$ and $B^*\bar{B}^*$ systems. 
These include unitarized chiral approaches~\cite{Cleven:2011gp,Cleven:2013sq,Hanhart:2015cua,Guo:2016bjq,Wang:2018jlv,Baru:2019xnh,Baru:2020ywb}, 
the Born-Oppenheimer analysis~\cite{Braaten:2014qka}, 
and lattice QCD simulations~\cite{Prelovsek:2019ywc,Hoffmann:2024hbz}. 
A precise determination of the $\delta_{Z_b}$ and $\delta_{Z_b^\prime}$ is therefore crucial for revealing the nature of the $Z_b$ and $Z_b^\prime$ mesons.

\begin{figure}[htb]
    \begin{center}
        \includegraphics[width=0.48\textwidth]{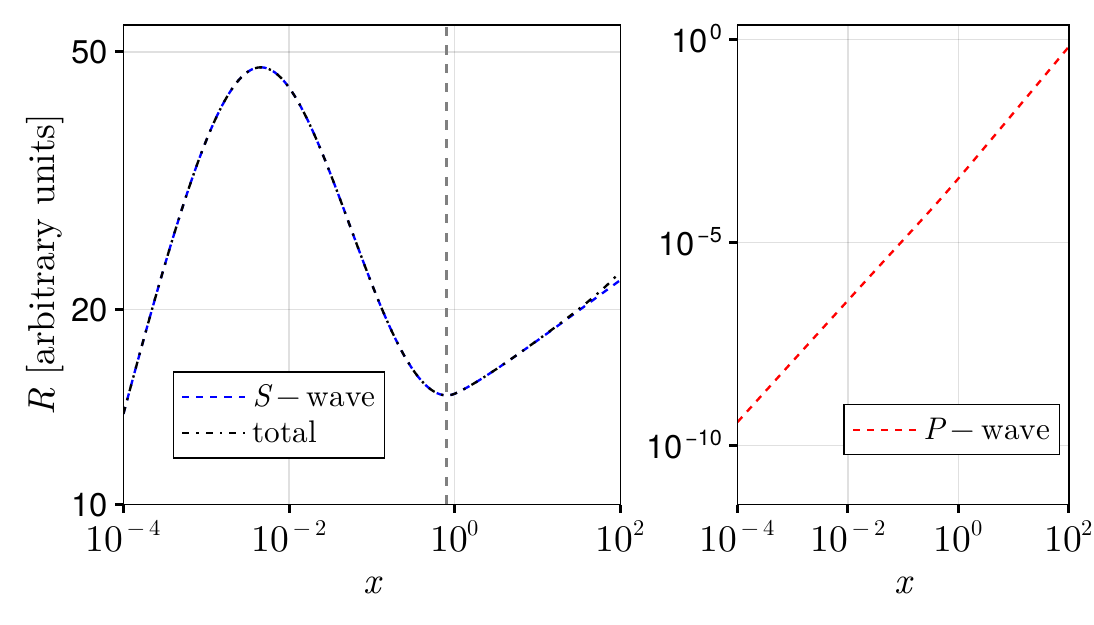}
    \end{center}
    \caption{\label{fig: ZbB}{The point production rates for the $Z_b B$ system with $(S, I)=(1,3/2)$ under the assumption that the $Z_b$ is a dimer state composed of $B^*\bar{B}$-$B\bar{B}^*$ with positive binding energy. The total point production rate exhibits a dip at $x=0.8022$, shown as the vertical dashed line in the left panel. }}
\end{figure}

The point production rates proposed in this work provide a novel and model-independent experimental method to determine the binding energies of the $Z_b$ and $Z_b^\prime$, benefiting from the negligible three-body effects in these systems~\cite{Lin:2017dbo}.
The total point production rate of the $(S,I)=(1,3/2)$ $Z_b B$ system with $\delta_{Z_b}=1\,{\rm MeV}$ is presented in Fig.~\ref{fig: ZbB}. 
In the low-energy region, the total rate exhibits a pronounced peak--dip structure dominated by the $S$-wave contribution, as the $P$-wave contribution is negligible.
Once the dip position $E_{\text{dip}}^{\text{exp}}$ is measured experimentally, the binding energy of the $Z_b$ can be extracted via the relation
\begin{equation}\label{eq: mass_Zb}
    \delta_{Z_b}=-\frac{E_{\text{dip}}^{\text{exp}}}{0.1983}\,.
\end{equation}
{Now we have established a linear relation between the dip position in the production rate and the binding energy of the corresponding dimer state, as given in Eq.~\eqref{eq: mass_Zb} for the $Z_b$. Owing to the sharp peak-dip structure, the experimental energy resolution directly impacts the precision of the extracted binding energy. A detailed analysis is presented in Appendix~\ref{app:resolution}. We find that this characteristic structure can be resolved only in experiments with an energy resolution better than approximately $0.4$~MeV, which is beyond the current capabilities of existing facilities. 
The proposed strategy is, however, promising for future experiments such as PANDA at FAIR~\cite{PANDA:2018zjt} and STCF~\cite{Achasov:2023gey}. Both are expected to perform high-precision energy scans in the charmonium mass region with an energy resolution of about $0.3$~MeV. This capability would enable a high-precision determination of the $X(3872)$ mass via short-range $XD$ production, potentially providing the first opportunity to implement the present method, as discussed below.
}
Eq.~\eqref{eq: mass_Zb} can be also applied to $Z_b^\prime$ with an accuracy better than $1\%$, owing to the closeness of $r_{Z_b} = 0.99148(4)$ and $r_{Z_b^\prime} = 1$.

Moreover, a similar strategy can be applied to extract the binding energy of $X(3872)$ from the measured
short-range production rate of the $XD$ system with $(S,I)=(1, 1/2)$. 
The corresponding partial-wave factor is $C_{1\frac12}=1/2$, and the RPP mass ratio is $r_X=M_D/M_{D^*}=0.92924$.
The dip position of the $XD$ system of $(S,I)=(1, 1/2)$ follows the same relation with the mass ratio as that of the $Z_b B$ system with $(S, I)=(1,3/2)$, since both systems share the same quantum number factor $C_{SI}$.
Therefore, the binding energy can be obtained immediately using
\begin{equation}\label{eq: mass_X}
    \delta_{X}=-\frac{E_{\text{dip}}^{\text{exp}}}{0.2126}\,.
\end{equation}
Generalizations of equations \eqref{eq: mass_Zb} and \eqref{eq: mass_X} for the finite range case are given in~\ref{app:ERC}. 
{Note also that the large coefficients in front of the dip positions in Eqs.~\eqref{eq: mass_Zb} and \eqref{eq: mass_X} amplify the uncertainties in the observed dip positions by roughly a factor of five in the resulting binding energies. For higher precision, a full line-shape fit to the observed distributions may be necessary.
} 

\section{Conclusion}
We have proposed a novel observable for measuring the binding energies of a wide class of near-threshold dimer states in a model-independent way. 
Here, we have focused on the discussion of the theoretical method. Some details on the experimental realization can be found in Ref.~\cite{Lin:2025tbm}.
In the absence of shallow three-body bound states and for small effective ranges, the point production rate of the dimer--spectator two-body system with in specific partial wave exhibits a universal peak--dip structure. 
In particular, the position of such dip is located at $\delta(x_{\rm dip} - 1)$, where $\delta$ is the dimer binding energy and $x_{\rm dip}$ is a universal, model-independent constant. This feature enables a precise extraction of their masses from experimental data,  provided that the line shape can be resolved with sufficient accuracy. For larger effective ranges, the binding energy can still be extracted by  performing a multi-parameter analysis of the line shape of the short-range production rate
calculated in effective field theory.
As concrete examples, we consider the $Z_b B$ and $Z_b^\prime B^*$ systems as well as the $XD$ system.
For all three cases, the binding energies of the near-threshold dimer can be extracted by measuring the dip position using 
Eqs.~\eqref{eq: mass_Zb} and \eqref{eq: mass_X}.
Prompt production has previously been used to study the properties of the $X(3872)$ at the Tevatron and the LHC \cite{Bauer:2004bc,Artoisenet:2009wk,ATLAS:2016kwu}. Our method provides another way to measure the binding energy of such threshold states. In this paper, we have treated the single-channel case. In the future it would be interesting to extend our results to systems with coupled channels. 


\textit{Acknowledgments.}~
HWH was supported by Deutsche Forschungsge-
meinschaft (DFG, German Research Foundation) under
Project ID 279384907 – SFB 1245 and by the German
BMFTR (Grant No. 05P24RDB). 
UGM was supported by the Chinese Academy of Sciences (CAS) President’s International Fellowship Initiative (PIFI) (Grant No. 2025PD0022), by the MKW NRW under the funding code NW21-024-A, and by the Deutsche Forschungsgemeinschaft (DFG,German Research Foundation) as part of the CRC 1639 NuMeriQS – project no. 511713970,
and by the European Research Council (ERC) under the European Union’s Horizon 2020 research and innovation programme (ERC AdG EXOTIC, grant agreement No. 101018170).

\appendix
\section{Finite-range corrections to point production}\label{app:ERC}
When the effective range expansion of the two-body amplitude is considered up to linear order in the effective range $\rho$,
\begin{equation}
    k\cot\delta_0(k)=-\gamma+\frac12 \rho\left(k^2+\gamma^2\right),
\end{equation}
an additional degree of freedom appears, expressed as the product of the effective range and the binding momentum, $\rho\gamma$~\cite{Phillips:1999hh,Hammer:2001gh,Bedaque:2002yg,Afnan:2003bs,Braaten:2004rn}. This quantity enters the integral equation of Eq.~\eqref{eq: ppdint} through the dimer propagator, and Eq.~\eqref{eq: ppdint} is modified to
\begin{align}\label{eq: inteq0NL1}
    &\Gamma_L(x,z)=g_L\left(2 \tilde{\mu} |\delta|\right)^{L/2}z^{L/2}+ (-1)^L C_{SI}\int_0^\infty \frac{dy}{2\pi\sqrt{z}}
    \notag\\
    &\phantom{x}\times \frac{(1+r)^2}{r\sqrt{2 r+1}} Q_L\left(\frac{(1+r)^2(y+z)-(1+2r)(x-1)}{2r(1+r) \sqrt{y z}}\right) \notag\\
    &\phantom{x} \times \left(\frac1{-{\rm sgn}(\delta)+\sqrt{1-x+y-i \epsilon}+\frac{\rho\gamma}{2}\left(x-y\right)}\right)\Gamma_L(x,y).
\end{align}
Using this equation, the effective-range effects for the short-range production can be investigated.
A major challenge in solving such an equation is that the range-corrected dimer propagator induces a spurious singularity that lies on the integration path in cases with positive effective range $\rho$. In the literature, these spurious poles are typically handled within a strict perturbative framework, where the two-body amplitude is expanded in the range parameter(s), and the effective-range corrections are included order by order according to a power-counting scheme~\cite{Hammer:2001gh,Bedaque:1998km,Ji:2011qg,Ji:2012nj,Vanasse:2013sda}. 
An alternative approach, designed for finite-volume calculations, was proposed more recently in Ref.~\cite{Ebert:2021epn}. 
{It should be noted that, in general, a three-body force is required to renormalize the ultraviolet divergence associated with range corrections, but 
this three-body force just restores the leading order renormalization condition in the three-body system. It does not add any new input parameters.}
For the systems considered here, the absence of a three-body force in the dimer--particle sector is a necessary prerequisite. 

We therefore solve the range-corrected integral equation by adopting a cutoff that is sufficiently low so that the spurious poles do not enter the integration contour. This treatment behaves well only for small values of $\rho\gamma$, where the higher spurious poles, located at $(x+4(1-\rho\gamma))/(\rho\gamma)^2$, remain far from the physical integration region and thus contribute negligibly, see Ref.~\cite{Lin:2025tbm} for the details.
\begin{figure*}[htbp]
    \begin{center}
        \includegraphics[width=0.6\textwidth]{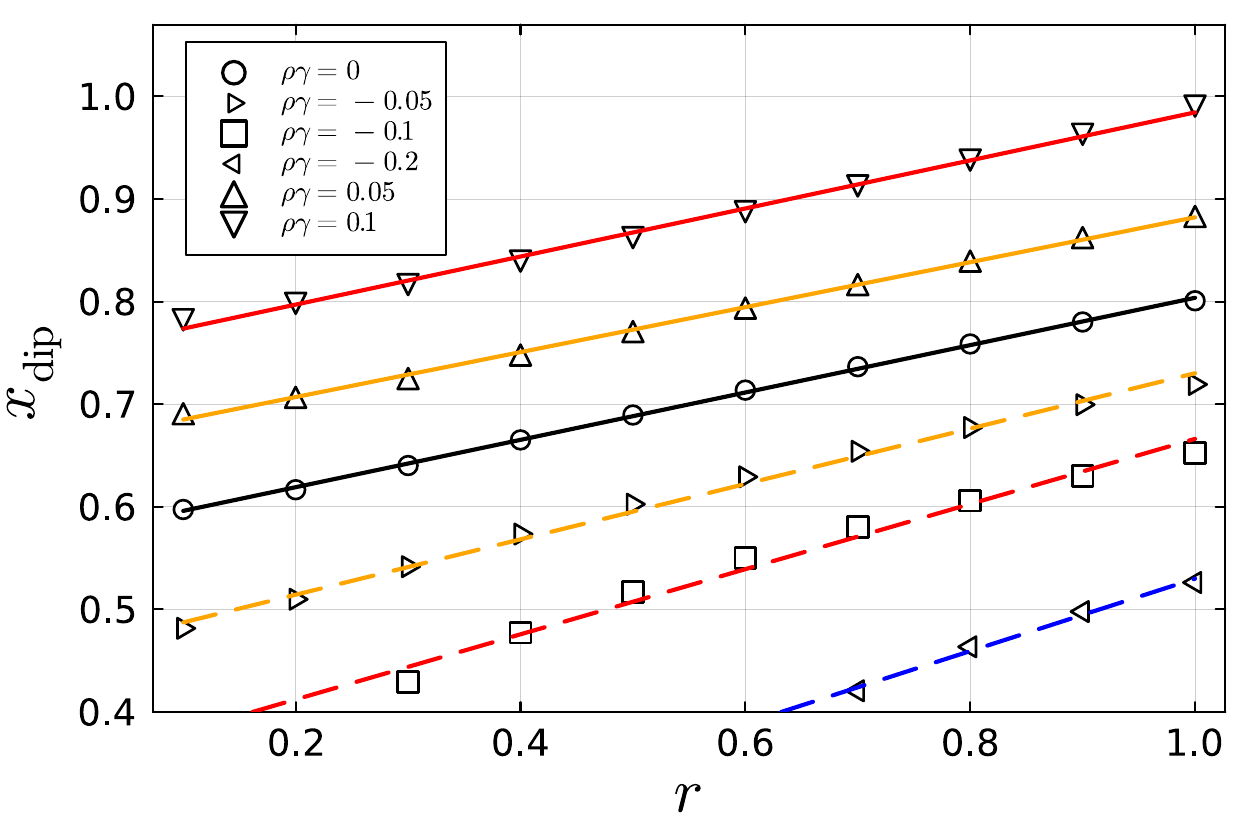}
    \end{center}
    \caption{\label{fig: erc1}{Dip position $x_{\text{dip}}$ versus mass ratio $r$ for various $\rho\gamma$. Symbols indicate explicit solutions of Eq.~\eqref{eq: inteq0NL1} while lines show linear fits.}}
\end{figure*}
Figure~\ref{fig: erc1} shows the numerical results for the correlation between the dip position $x_{\text{dip}}$ and the mass ratio $r$ at different values of $\rho\gamma$, with $C_{SI}=1/2$ taken without loss of generality. 
The corresponding linear fits are summarized in Table~\ref{Tab: linear}. 
For $-0.1 \leq \rho\gamma \leq 0.1$, a pronounced peak-dip structure is observed, and the dip position exhibits an approximately linear dependence on the mass ratio.
This also enables a direct extraction of the binding energy from the measured dip position, using the relations given in the third and fourth columns of Table~\ref{Tab: linear} for the $Z_b$ and $X(3872)$ cases, respectively.

For larger values of $|\rho\gamma|$, however, the peak-dip structure is washed out, requiring a more sophisticated analysis of the line shape to extract both the finite-range parameter and the binding energy. As shown in Fig.~\ref{fig: erc1}, for $\rho\gamma=-0.2$ the peak-dip structure survives only at large mass ratios $r\gtrsim 0.6$. However, for the examples $Z_b B$, $Z_b^\prime B^*$, and $XD$ considered in the main text and other potential applications of the method, the mass ratios are larger than 0.9.
Thus, the effective range plays a crucial role in the short-range production of the dimer-particle system. 
The linear extraction of the binding energy of a near-threshold state with the proposed method is valid only within a narrow window near the zero-range limit. 

For larger effective ranges, $|\rho\gamma|\gtrsim 0.1$, a multi-parameter analysis of the line shape is required to determine both the binding energy of the near-threshold state and the effective range of the corresponding two-body subsystem. Within a short-range effective field theory, the dependence of the line shape on the effective range can be computed reliably for $|\gamma\rho| \lesssim 0.3$. For effective ranges $|\rho\gamma|\gtrsim 0.3$, the state moves away from the threshold and our method is not applicable.
\begin{table}[htbp]
    \centering
    \renewcommand\arraystretch{1.2}
    \caption{Explicit linear relations of the dip position $x_{\text{dip}}$ versus mass ratio $r$ for various $\rho\gamma$.
        The corresponding binding-energy relations for $Z_b$ and $X(3872)$ are given in the third and fourth columns, respectively.\label{Tab: linear}}
    \begin{tabular}{c| c c c}
        \hline
        \hline
        $\rho\gamma$ & Linear relation & \shortstack{Variation \\ of Eq.~\eqref{eq: mass_Zb}} & \shortstack{Variation \\ of Eq.~\eqref{eq: mass_X}}  \\
        \hline
        $0$  & $0.2308 r +0.5729$ & $-{E_{\text{dip}}^{\text{exp}}}/{0.1983}$ & $-{E_{\text{dip}}^{\text{exp}}}/{0.2126}$  \\
        $-0.05$  & $0.2699 r +0.4603$ & $-{E_{\text{dip}}^{\text{exp}}}/{0.2721}$ & $-{E_{\text{dip}}^{\text{exp}}}/{0.2889}$  \\
        $-0.1$  & $0.3175 r +0.3487$ & $-{E_{\text{dip}}^{\text{exp}}}/{0.3365}$ & $-{E_{\text{dip}}^{\text{exp}}}/{0.3563}$  \\
        $0.05$  & $0.2191 r +0.6631$ & $-{E_{\text{dip}}^{\text{exp}}}/{0.1197}$ & $-{E_{\text{dip}}^{\text{exp}}}/{0.1333}$  \\
        $0.1$  & $0.2342 r +0.7502$  & $-{E_{\text{dip}}^{\text{exp}}}/{0.0176}$ & $-{E_{\text{dip}}^{\text{exp}}}/{0.0322}$  \\
        \hline
        \hline
    \end{tabular}
\end{table}

{
\section{Energy resolution effect}\label{app:resolution}
We briefly examine the uncertainty in determining the dip position arising from the finite energy resolution of the detector. In experiments, the measured distribution corresponds to the convolution of the true distribution with the detector resolution function. This effect becomes particularly important for sharp structures, as in the present case. To quantify the resulting distortion, we simulate the measured line shape by evaluating the convolution integral assuming a Gaussian resolution function,
\begin{equation}
    R_{\text{obs}}(x)=\int_{-\infty}^{\infty}d x^\prime R_{\text{true}}(x^\prime)G(x-x^\prime), 
\end{equation}
where the resolution function $G$ is taken to be
\begin{equation}
    G(x-x^\prime)=\frac1{\sqrt{2\pi}\sigma}\exp \left(-\frac{(x-x^\prime)^2}{2\sigma^2}\right),
\end{equation}
with $\sigma$ being the detector energy resolution.
Figure~\ref{fig: Eres} shows the smeared line shapes (blue solid curves) for the dimer--particle production rate in the $Z_b B$ channel, compared with the true distributions (black dashed curves), for three different energy resolutions, $\sigma=0.1$, $0.3$, and $0.4$~MeV. A visible shift of the dip position emerges once the detector resolution exceeds $0.1$~MeV. The shift is estimated to be about $0.2$~MeV for $\sigma=0.3$~MeV and $1.2$~MeV for $\sigma=0.4$~MeV. This effect introduces an additional uncertainty in the extracted binding energy when Eqs.~\eqref{eq: mass_Zb} and \eqref{eq: mass_X} are employed.

In practice, the experimentally measured dip position should first be corrected by subtracting the corresponding three-bottom-meson mass threshold before being inserted into Eqs.~\eqref{eq: mass_Zb} and \eqref{eq: mass_X}. Furthermore, when $\sigma \ge 0.5$~MeV, the characteristic peak-dip structure becomes indistinguishable, rendering the extraction strategy ineffective. This indicates that the proposed method is applicable only at experimental facilities with an energy resolution better than about $0.4$~MeV.
\begin{figure*}[htbp]
    \begin{center}
        \includegraphics[width=0.85\textwidth]{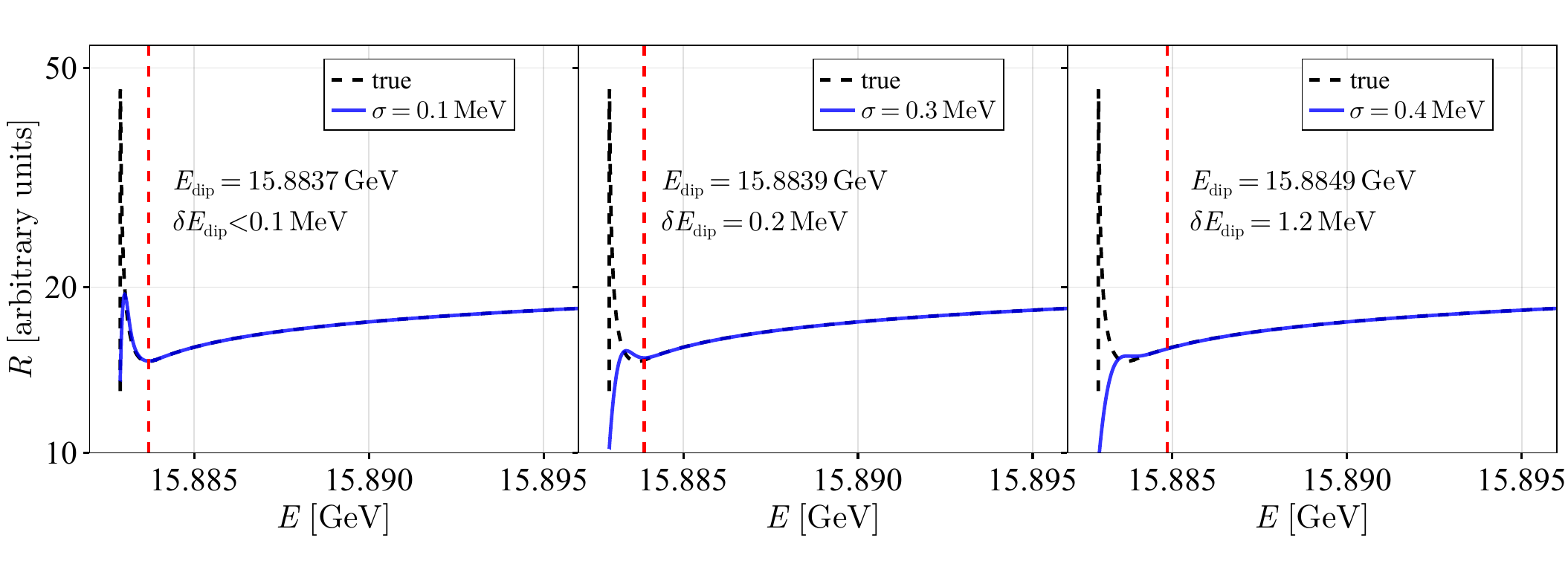}
    \end{center}
    \caption{\label{fig: Eres}{
            Dip position shifts induced by the finite detector resolution. The blue solid curves show the smeared line shapes, compared with the true production rate in the $Z_b B$ channel (black dashed curves). The quantity $\delta E_{\text{dip}}$ denotes the corresponding shift of the dip position due to the finite detector resolution. The extracted $E_{\text{dip}}$ should be measured relative to the corresponding three-bottom-meson threshold before being inserted into Eqs.~\eqref{eq: mass_Zb} and \eqref{eq: mass_X}.}}
\end{figure*}
}

\section*{References}

\bibliographystyle{elsarticle-num}
\bibliography{refs}

\begin{thebibliography}{10}
\expandafter\ifx\csname url\endcsname\relax
  \def\url#1{\texttt{#1}}\fi
\expandafter\ifx\csname urlprefix\endcsname\relax\def\urlprefix{URL }\fi
\expandafter\ifx\csname href\endcsname\relax
  \def\href#1#2{#2} \def\path#1{#1}\fi

\bibitem{Hosaka:2016pey}
A.~Hosaka, T.~Iijima, K.~Miyabayashi, Y.~Sakai, S.~Yasui, {Exotic hadrons with
  heavy flavors: X, Y, Z, and related states}, PTEP 2016~(6) (2016) 062C01.
\newblock \href {http://arxiv.org/abs/1603.09229} {\path{arXiv:1603.09229}},
  \href {https://doi.org/10.1093/ptep/ptw045} {\path{doi:10.1093/ptep/ptw045}}.

\bibitem{Esposito:2016noz}
A.~Esposito, A.~Pilloni, A.~D. Polosa, {Multiquark Resonances}, Phys. Rept. 668
  (2017) 1--97.
\newblock \href {http://arxiv.org/abs/1611.07920} {\path{arXiv:1611.07920}},
  \href {https://doi.org/10.1016/j.physrep.2016.11.002}
  {\path{doi:10.1016/j.physrep.2016.11.002}}.

\bibitem{Guo:2017jvc}
F.-K. Guo, C.~Hanhart, U.-G. Mei{\ss}ner, Q.~Wang, Q.~Zhao, B.-S. Zou,
  {Hadronic molecules}, Rev. Mod. Phys. 90~(1) (2018) 015004, [Erratum:
  Rev.Mod.Phys. 94, 029901 (2022)].
\newblock \href {http://arxiv.org/abs/1705.00141} {\path{arXiv:1705.00141}},
  \href {https://doi.org/10.1103/RevModPhys.90.015004}
  {\path{doi:10.1103/RevModPhys.90.015004}}.

\bibitem{Olsen:2017bmm}
S.~L. Olsen, T.~Skwarnicki, D.~Zieminska, {Nonstandard heavy mesons and
  baryons: Experimental evidence}, Rev. Mod. Phys. 90~(1) (2018) 015003.
\newblock \href {http://arxiv.org/abs/1708.04012} {\path{arXiv:1708.04012}},
  \href {https://doi.org/10.1103/RevModPhys.90.015003}
  {\path{doi:10.1103/RevModPhys.90.015003}}.

\bibitem{Karliner:2017qhf}
M.~Karliner, J.~L. Rosner, T.~Skwarnicki, {Multiquark States}, Ann. Rev. Nucl.
  Part. Sci. 68 (2018) 17--44.
\newblock \href {http://arxiv.org/abs/1711.10626} {\path{arXiv:1711.10626}},
  \href {https://doi.org/10.1146/annurev-nucl-101917-020902}
  {\path{doi:10.1146/annurev-nucl-101917-020902}}.

\bibitem{Kalashnikova:2018vkv}
Y.~S. Kalashnikova, A.~V. Nefediev, {X(3872) in the molecular model}, Phys.
  Usp. 62~(6) (2019) 568--595.
\newblock \href {http://arxiv.org/abs/1811.01324} {\path{arXiv:1811.01324}},
  \href {https://doi.org/10.3367/UFNe.2018.08.038411}
  {\path{doi:10.3367/UFNe.2018.08.038411}}.

\bibitem{Brambilla:2019esw}
N.~Brambilla, S.~Eidelman, C.~Hanhart, A.~Nefediev, C.-P. Shen, C.~E. Thomas,
  A.~Vairo, C.-Z. Yuan, {The $XYZ$ states: experimental and theoretical status
  and perspectives}, Phys. Rept. 873 (2020) 1--154.
\newblock \href {http://arxiv.org/abs/1907.07583} {\path{arXiv:1907.07583}},
  \href {https://doi.org/10.1016/j.physrep.2020.05.001}
  {\path{doi:10.1016/j.physrep.2020.05.001}}.

\bibitem{Meng:2022ozq}
L.~Meng, B.~Wang, G.-J. Wang, S.-L. Zhu, {Chiral perturbation theory for heavy
  hadrons and chiral effective field theory for heavy hadronic molecules},
  Phys. Rept. 1019 (2023) 1--149.
\newblock \href {http://arxiv.org/abs/2204.08716} {\path{arXiv:2204.08716}},
  \href {https://doi.org/10.1016/j.physrep.2023.04.003}
  {\path{doi:10.1016/j.physrep.2023.04.003}}.

\bibitem{Liu:2024uxn}
M.-Z. Liu, Y.-W. Pan, Z.-W. Liu, T.-W. Wu, J.-X. Lu, L.-S. Geng, {Three ways to
  decipher the nature of exotic hadrons: Multiplets, three-body hadronic
  molecules, and correlation functions}, Phys. Rept. 1108 (2025) 1--108.
\newblock \href {http://arxiv.org/abs/2404.06399} {\path{arXiv:2404.06399}},
  \href {https://doi.org/10.1016/j.physrep.2024.12.001}
  {\path{doi:10.1016/j.physrep.2024.12.001}}.

\bibitem{Chen:2024eaq}
J.-H. Chen, J.~Chen, F.-K. Guo, Y.-G. Ma, C.-P. Shen, Q.-Y. Shou, Q.~Shou,
  Q.~Wang, J.-J. Wu, B.-S. Zou, {Production of exotic hadrons in pp and nuclear
  collisions}, Nucl. Sci. Tech. 36~(4) (2025) 55.
\newblock \href {http://arxiv.org/abs/2411.18257} {\path{arXiv:2411.18257}},
  \href {https://doi.org/10.1007/s41365-025-01664-w}
  {\path{doi:10.1007/s41365-025-01664-w}}.

\bibitem{ParticleDataGroup:2024cfk}
S.~Navas, et~al., {Review of particle physics}, Phys. Rev. D 110~(3) (2024)
  030001.
\newblock \href {https://doi.org/10.1103/PhysRevD.110.030001}
  {\path{doi:10.1103/PhysRevD.110.030001}}.

\bibitem{Guo:2019qcn}
F.-K. Guo, {Novel Method for Precisely Measuring the $X(3872)$ Mass}, Phys.
  Rev. Lett. 122~(20) (2019) 202002.
\newblock \href {http://arxiv.org/abs/1902.11221} {\path{arXiv:1902.11221}},
  \href {https://doi.org/10.1103/PhysRevLett.122.202002}
  {\path{doi:10.1103/PhysRevLett.122.202002}}.

\bibitem{Mehen:1999nd}
T.~Mehen, I.~W. Stewart, M.~B. Wise, {Conformal invariance for nonrelativistic
  field theory}, Phys. Lett. B 474 (2000) 145--152.
\newblock \href {http://arxiv.org/abs/hep-th/9910025}
  {\path{arXiv:hep-th/9910025}}, \href
  {https://doi.org/10.1016/S0370-2693(00)00006-X}
  {\path{doi:10.1016/S0370-2693(00)00006-X}}.

\bibitem{Braaten:2004rn}
E.~Braaten, H.~W. Hammer, {Universality in few-body systems with large
  scattering length}, Phys. Rept. 428 (2006) 259--390.
\newblock \href {http://arxiv.org/abs/cond-mat/0410417}
  {\path{arXiv:cond-mat/0410417}}, \href
  {https://doi.org/10.1016/j.physrep.2006.03.001}
  {\path{doi:10.1016/j.physrep.2006.03.001}}.

\bibitem{Nishida:2007pj}
Y.~Nishida, D.~T. Son, {Nonrelativistic conformal field theories}, Phys. Rev. D
  76 (2007) 086004.
\newblock \href {http://arxiv.org/abs/0706.3746} {\path{arXiv:0706.3746}},
  \href {https://doi.org/10.1103/PhysRevD.76.086004}
  {\path{doi:10.1103/PhysRevD.76.086004}}.

\bibitem{Hammer:2021zxb}
H.-W. Hammer, D.~T. Son, {Unnuclear physics}, Proc. Nat. Acad. Sci. 118 (2021)
  e2108716118.
\newblock \href {http://arxiv.org/abs/2103.12610} {\path{arXiv:2103.12610}},
  \href {https://doi.org/10.1073/pnas.2108716118}
  {\path{doi:10.1073/pnas.2108716118}}.

\bibitem{Braaten:2021iot}
E.~Braaten, H.-W. Hammer, {Interpretation of Neutral Charm Mesons near
  Threshold as Unparticles}, Phys. Rev. Lett. 128~(3) (2022) 032002.
\newblock \href {http://arxiv.org/abs/2107.02831} {\path{arXiv:2107.02831}},
  \href {https://doi.org/10.1103/PhysRevLett.128.032002}
  {\path{doi:10.1103/PhysRevLett.128.032002}}.

\bibitem{Braaten:2023acw}
E.~Braaten, H.-W. Hammer, {Point production of a nonrelativistic unparticle
  recoiling against a particle}, Phys. Rev. D 107~(3) (2023) 034017.
\newblock \href {http://arxiv.org/abs/2301.04399} {\path{arXiv:2301.04399}},
  \href {https://doi.org/10.1103/PhysRevD.107.034017}
  {\path{doi:10.1103/PhysRevD.107.034017}}.

\bibitem{Wilbring:2016bda}
E.~Wilbring, {Efimov Effect in Pionless Effective Field Theory and its
  Application to Hadronic Molecules}, PhD thesis, Rheinische
  Friedrich-Wilhelms-Universität Bonn (2016)\href
  {https://doi.org/https://hdl.handle.net/20.500.11811/6714}
  {\path{doi:https://hdl.handle.net/20.500.11811/6714}}.

\bibitem{Lin:2017dbo}
Y.-H. Lin, E.~Wilbring, H.-L. Fu, H.-W. Hammer, U.-G. Mei{\ss}ner, {Three-body
  universality in the B meson sector}, J. Phys. G 52~(10) (2025) 105005.
\newblock \href {http://arxiv.org/abs/1705.06176} {\path{arXiv:1705.06176}},
  \href {https://doi.org/10.1088/1361-6471/ae06be}
  {\path{doi:10.1088/1361-6471/ae06be}}.

\bibitem{Fu:2025joa}
H.-L. Fu, Y.-H. Lin, F.-K. Guo, H.-W. Hammer, U.-G. Mei{\ss}ner, A.~Rusetsky,
  X.~Zhang, {Exploring Efimov states in D$^{*}$D$^{*}$D$^{*}$ and
  DD$^{*}$D$^{*}$ three-body systems}, JHEP 07 (2025) 081.
\newblock \href {http://arxiv.org/abs/2503.19709} {\path{arXiv:2503.19709}},
  \href {https://doi.org/10.1007/JHEP07(2025)081}
  {\path{doi:10.1007/JHEP07(2025)081}}.

\bibitem{Bedaque:1998kg}
P.~F. Bedaque, H.~W. Hammer, U.~van Kolck, {Renormalization of the three-body
  system with short range interactions}, Phys. Rev. Lett. 82 (1999) 463--467.
\newblock \href {http://arxiv.org/abs/nucl-th/9809025}
  {\path{arXiv:nucl-th/9809025}}, \href
  {https://doi.org/10.1103/PhysRevLett.82.463}
  {\path{doi:10.1103/PhysRevLett.82.463}}.

\bibitem{Bedaque:1998km}
P.~F. Bedaque, H.~W. Hammer, U.~van Kolck, {The Three boson system with short
  range interactions}, Nucl. Phys. A 646 (1999) 444--466.
\newblock \href {http://arxiv.org/abs/nucl-th/9811046}
  {\path{arXiv:nucl-th/9811046}}, \href
  {https://doi.org/10.1016/S0375-9474(98)00650-2}
  {\path{doi:10.1016/S0375-9474(98)00650-2}}.

\bibitem{Bedaque:1999ve}
P.~F. Bedaque, H.~W. Hammer, U.~van Kolck, {Effective theory of the triton},
  Nucl. Phys. A 676 (2000) 357--370.
\newblock \href {http://arxiv.org/abs/nucl-th/9906032}
  {\path{arXiv:nucl-th/9906032}}, \href
  {https://doi.org/10.1016/S0375-9474(00)00205-0}
  {\path{doi:10.1016/S0375-9474(00)00205-0}}.

\bibitem{Efimov:1970zz}
V.~Efimov, {Energy levels arising form the resonant two-body forces in a
  three-body system}, Phys. Lett. B 33 (1970) 563--564.
\newblock \href {https://doi.org/10.1016/0370-2693(70)90349-7}
  {\path{doi:10.1016/0370-2693(70)90349-7}}.

\bibitem{Bedaque:1997qi}
P.~F. Bedaque, U.~van Kolck, {Nucleon deuteron scattering from an effective
  field theory}, Phys. Lett. B 428 (1998) 221--226.
\newblock \href {http://arxiv.org/abs/nucl-th/9710073}
  {\path{arXiv:nucl-th/9710073}}, \href
  {https://doi.org/10.1016/S0370-2693(98)00430-4}
  {\path{doi:10.1016/S0370-2693(98)00430-4}}.

\bibitem{Bedaque:1998mb}
P.~F. Bedaque, H.~W. Hammer, U.~van Kolck, {Effective theory for neutron
  deuteron scattering: Energy dependence}, Phys. Rev. C 58 (1998) R641--R644.
\newblock \href {http://arxiv.org/abs/nucl-th/9802057}
  {\path{arXiv:nucl-th/9802057}}, \href
  {https://doi.org/10.1103/PhysRevC.58.R641}
  {\path{doi:10.1103/PhysRevC.58.R641}}.

\bibitem{Hammer:2010kp}
H.-W. Hammer, L.~Platter, {Efimov States in Nuclear and Particle Physics}, Ann.
  Rev. Nucl. Part. Sci. 60 (2010) 207--236.
\newblock \href {http://arxiv.org/abs/1001.1981} {\path{arXiv:1001.1981}},
  \href {https://doi.org/10.1146/annurev.nucl.012809.104439}
  {\path{doi:10.1146/annurev.nucl.012809.104439}}.

\bibitem{Epelbaum:2016ffd}
E.~Epelbaum, J.~Gegelia, U.-G. Mei{\ss}ner, D.-L. Yao, {Renormalization of the
  three-boson system with short-range interactions revisited}, Eur. Phys. J. A
  53~(5) (2017) 98.
\newblock \href {http://arxiv.org/abs/1611.06040} {\path{arXiv:1611.06040}},
  \href {https://doi.org/10.1140/epja/i2017-12288-3}
  {\path{doi:10.1140/epja/i2017-12288-3}}.

\bibitem{Cleven:2011gp}
M.~Cleven, F.-K. Guo, C.~Hanhart, U.-G. Meissner, {Bound state nature of the
  exotic $Z_b$ states}, Eur. Phys. J. A 47 (2011) 120.
\newblock \href {http://arxiv.org/abs/1107.0254} {\path{arXiv:1107.0254}},
  \href {https://doi.org/10.1140/epja/i2011-11120-6}
  {\path{doi:10.1140/epja/i2011-11120-6}}.

\bibitem{Cleven:2013sq}
M.~Cleven, Q.~Wang, F.-K. Guo, C.~Hanhart, U.-G. Meissner, Q.~Zhao, {Confirming
  the molecular nature of the $Z_b(10610)$ and the $Z_b(10650)$}, Phys. Rev. D
  87~(7) (2013) 074006.
\newblock \href {http://arxiv.org/abs/1301.6461} {\path{arXiv:1301.6461}},
  \href {https://doi.org/10.1103/PhysRevD.87.074006}
  {\path{doi:10.1103/PhysRevD.87.074006}}.

\bibitem{Hanhart:2015cua}
C.~Hanhart, Y.~S. Kalashnikova, P.~Matuschek, R.~V. Mizuk, A.~V. Nefediev,
  Q.~Wang, {Practical Parametrization for Line Shapes of Near-Threshold
  States}, Phys. Rev. Lett. 115~(20) (2015) 202001.
\newblock \href {http://arxiv.org/abs/1507.00382} {\path{arXiv:1507.00382}},
  \href {https://doi.org/10.1103/PhysRevLett.115.202001}
  {\path{doi:10.1103/PhysRevLett.115.202001}}.

\bibitem{Guo:2016bjq}
F.~K. Guo, C.~Hanhart, Y.~S. Kalashnikova, P.~Matuschek, R.~V. Mizuk, A.~V.
  Nefediev, Q.~Wang, J.~L. Wynen, {Interplay of quark and meson degrees of
  freedom in near-threshold states: A practical parametrization for line
  shapes}, Phys. Rev. D 93~(7) (2016) 074031.
\newblock \href {http://arxiv.org/abs/1602.00940} {\path{arXiv:1602.00940}},
  \href {https://doi.org/10.1103/PhysRevD.93.074031}
  {\path{doi:10.1103/PhysRevD.93.074031}}.

\bibitem{Wang:2018jlv}
Q.~Wang, V.~Baru, A.~A. Filin, C.~Hanhart, A.~V. Nefediev, J.~L. Wynen, {Line
  shapes of the $Z_b(10610)$ and $Z_b(10650)$ in the elastic and inelastic
  channels revisited}, Phys. Rev. D 98~(7) (2018) 074023.
\newblock \href {http://arxiv.org/abs/1805.07453} {\path{arXiv:1805.07453}},
  \href {https://doi.org/10.1103/PhysRevD.98.074023}
  {\path{doi:10.1103/PhysRevD.98.074023}}.

\bibitem{Baru:2019xnh}
V.~Baru, E.~Epelbaum, A.~A. Filin, C.~Hanhart, A.~V. Nefediev, Q.~Wang, {Spin
  partners $W_{bJ}$ from the line shapes of the $Z_b(10610)$ and $Z_b(10650)$},
  Phys. Rev. D 99~(9) (2019) 094013.
\newblock \href {http://arxiv.org/abs/1901.10319} {\path{arXiv:1901.10319}},
  \href {https://doi.org/10.1103/PhysRevD.99.094013}
  {\path{doi:10.1103/PhysRevD.99.094013}}.

\bibitem{Baru:2020ywb}
V.~Baru, E.~Epelbaum, A.~A. Filin, C.~Hanhart, R.~V. Mizuk, A.~V. Nefediev,
  S.~Ropertz, {Insights into $Z_b(10610)$ and $Z_b(10650)$ from dipion
  transitions from $\Upsilon(10860)$}, Phys. Rev. D 103~(3) (2021) 034016.
\newblock \href {http://arxiv.org/abs/2012.05034} {\path{arXiv:2012.05034}},
  \href {https://doi.org/10.1103/PhysRevD.103.034016}
  {\path{doi:10.1103/PhysRevD.103.034016}}.

\bibitem{Braaten:2014qka}
E.~Braaten, C.~Langmack, D.~H. Smith, {Born-Oppenheimer Approximation for the
  XYZ Mesons}, Phys. Rev. D 90~(1) (2014) 014044.
\newblock \href {http://arxiv.org/abs/1402.0438} {\path{arXiv:1402.0438}},
  \href {https://doi.org/10.1103/PhysRevD.90.014044}
  {\path{doi:10.1103/PhysRevD.90.014044}}.

\bibitem{Prelovsek:2019ywc}
S.~Prelovsek, H.~Bahtiyar, J.~Petkovic, {Zb tetraquark channel from lattice QCD
  and Born-Oppenheimer approximation}, Phys. Lett. B 805 (2020) 135467.
\newblock \href {http://arxiv.org/abs/1912.02656} {\path{arXiv:1912.02656}},
  \href {https://doi.org/10.1016/j.physletb.2020.135467}
  {\path{doi:10.1016/j.physletb.2020.135467}}.

\bibitem{Hoffmann:2024hbz}
J.~Hoffmann, M.~Wagner, {Prediction of an I(JP)=0(1-)
  b{\textasciimacron}b{\textasciimacron}ud tetraquark resonance close to the
  B*B* threshold using lattice QCD potentials}, Phys. Rev. D 111~(5) (2025)
  054507.
\newblock \href {http://arxiv.org/abs/2412.06607} {\path{arXiv:2412.06607}},
  \href {https://doi.org/10.1103/PhysRevD.111.054507}
  {\path{doi:10.1103/PhysRevD.111.054507}}.

\bibitem{PANDA:2018zjt}
G.~Barucca, et~al., {Precision resonance energy scans with the PANDA experiment
  at FAIR: Sensitivity study for width and line-shape measurements of the
  X(3872)}, Eur. Phys. J. A 55~(3) (2019) 42.
\newblock \href {http://arxiv.org/abs/1812.05132} {\path{arXiv:1812.05132}},
  \href {https://doi.org/10.1140/epja/i2019-12718-2}
  {\path{doi:10.1140/epja/i2019-12718-2}}.

\bibitem{Achasov:2023gey}
M.~Achasov, et~al., {STCF conceptual design report (Volume 1): Physics {\&}
  detector}, Front. Phys. (Beijing) 19~(1) (2024) 14701.
\newblock \href {http://arxiv.org/abs/2303.15790} {\path{arXiv:2303.15790}},
  \href {https://doi.org/10.1007/s11467-023-1333-z}
  {\path{doi:10.1007/s11467-023-1333-z}}.

\bibitem{Lin:2025tbm}
Y.-H. Lin, H.-W. Hammer, U.-G. Mei{\ss}ner, {Short-range production of three
  bottom mesons}, JHEP 04 (2026) 092.
\newblock \href {http://arxiv.org/abs/2511.22590} {\path{arXiv:2511.22590}},
  \href {https://doi.org/10.1007/JHEP04(2026)092}
  {\path{doi:10.1007/JHEP04(2026)092}}.

\bibitem{Bauer:2004bc}
G.~Bauer, {The $X(3872) $ at CDF II}, Int. J. Mod. Phys. A 20 (2005)
  3765--3767.
\newblock \href {http://arxiv.org/abs/hep-ex/0409052}
  {\path{arXiv:hep-ex/0409052}}, \href
  {https://doi.org/10.1142/S0217751X05027552}
  {\path{doi:10.1142/S0217751X05027552}}.

\bibitem{Artoisenet:2009wk}
P.~Artoisenet, E.~Braaten, {Production of the X(3872) at the Tevatron and the
  LHC}, Phys. Rev. D 81 (2010) 114018.
\newblock \href {http://arxiv.org/abs/0911.2016} {\path{arXiv:0911.2016}},
  \href {https://doi.org/10.1103/PhysRevD.81.114018}
  {\path{doi:10.1103/PhysRevD.81.114018}}.

\bibitem{ATLAS:2016kwu}
M.~Aaboud, et~al., {Measurements of $\psi(2S)$ and $X(3872) \to
  J/\psi\pi^+\pi^-$ production in $pp$ collisions at $\sqrt{s} = 8$ TeV with
  the ATLAS detector}, JHEP 01 (2017) 117.
\newblock \href {http://arxiv.org/abs/1610.09303} {\path{arXiv:1610.09303}},
  \href {https://doi.org/10.1007/JHEP01(2017)117}
  {\path{doi:10.1007/JHEP01(2017)117}}.

\bibitem{Phillips:1999hh}
D.~R. Phillips, G.~Rupak, M.~J. Savage, {Improving the convergence of N N
  effective field theory}, Phys. Lett. B 473 (2000) 209--218.
\newblock \href {http://arxiv.org/abs/nucl-th/9908054}
  {\path{arXiv:nucl-th/9908054}}, \href
  {https://doi.org/10.1016/S0370-2693(99)01496-3}
  {\path{doi:10.1016/S0370-2693(99)01496-3}}.

\bibitem{Hammer:2001gh}
H.~W. Hammer, T.~Mehen, {Range corrections to doublet S wave neutron deuteron
  scattering}, Phys. Lett. B 516 (2001) 353--361.
\newblock \href {http://arxiv.org/abs/nucl-th/0105072}
  {\path{arXiv:nucl-th/0105072}}, \href
  {https://doi.org/10.1016/S0370-2693(01)00918-2}
  {\path{doi:10.1016/S0370-2693(01)00918-2}}.

\bibitem{Bedaque:2002yg}
P.~F. Bedaque, G.~Rupak, H.~W. Griesshammer, H.-W. Hammer, {Low-energy
  expansion in the three-body system to all orders and the triton channel},
  Nucl. Phys. A 714 (2003) 589--610.
\newblock \href {http://arxiv.org/abs/nucl-th/0207034}
  {\path{arXiv:nucl-th/0207034}}, \href
  {https://doi.org/10.1016/S0375-9474(02)01402-1}
  {\path{doi:10.1016/S0375-9474(02)01402-1}}.

\bibitem{Afnan:2003bs}
I.~R. Afnan, D.~R. Phillips, {The Three body problem with short range forces:
  Renormalized equations and regulator independent results}, Phys. Rev. C 69
  (2004) 034010.
\newblock \href {http://arxiv.org/abs/nucl-th/0312021}
  {\path{arXiv:nucl-th/0312021}}, \href
  {https://doi.org/10.1103/PhysRevC.69.034010}
  {\path{doi:10.1103/PhysRevC.69.034010}}.

\bibitem{Ji:2011qg}
C.~Ji, D.~R. Phillips, L.~Platter, {The three-boson system at next-to-leading
  order in an effective field theory for systems with a large scattering
  length}, Annals Phys. 327 (2012) 1803--1824.
\newblock \href {http://arxiv.org/abs/1106.3837} {\path{arXiv:1106.3837}},
  \href {https://doi.org/10.1016/j.aop.2012.02.001}
  {\path{doi:10.1016/j.aop.2012.02.001}}.

\bibitem{Ji:2012nj}
C.~Ji, D.~R. Phillips, {Effective Field Theory Analysis of Three-Boson Systems
  at Next-To-Next-To-Leading Order}, Few Body Syst. 54 (2013) 2317--2355.
\newblock \href {http://arxiv.org/abs/1212.1845} {\path{arXiv:1212.1845}},
  \href {https://doi.org/10.1007/s00601-013-0710-5}
  {\path{doi:10.1007/s00601-013-0710-5}}.

\bibitem{Vanasse:2013sda}
J.~Vanasse, {Fully Perturbative Calculation of $nd$ Scattering to
  Next-to-next-to-leading-order}, Phys. Rev. C 88~(4) (2013) 044001.
\newblock \href {http://arxiv.org/abs/1305.0283} {\path{arXiv:1305.0283}},
  \href {https://doi.org/10.1103/PhysRevC.88.044001}
  {\path{doi:10.1103/PhysRevC.88.044001}}.

\bibitem{Ebert:2021epn}
M.~Ebert, H.~W. Hammer, A.~Rusetsky, {An alternative scheme for effective range
  corrections in pionless EFT}, Eur. Phys. J. A 57~(12) (2021) 332.
\newblock \href {http://arxiv.org/abs/2109.11982} {\path{arXiv:2109.11982}},
  \href {https://doi.org/10.1140/epja/s10050-021-00637-y}
  {\path{doi:10.1140/epja/s10050-021-00637-y}}.

\end{thebibliography}

\end{document}